\documentclass[showpacs]{revtex4}

\usepackage{graphicx}
\usepackage{dcolumn}
\usepackage{amsmath}
\usepackage{color}

\makeatletter
\def\btt#1{\texttt{\@backslashchar#1}}
\DeclareRobustCommand\bblash{\btt{\@backslashchar}} \makeatother

\begin{document}
\title{Noncommutative geometry inspired Einstein-Gauss-Bonnet black holes}
\author{Sushant G. Ghosh}
\email{sgghosh@gmail.com,sghosh2@jmi.ac.in}
\affiliation{Centre for Theoretical Physics, Jamia Millia Islamia,  New Delhi 110025, India}
\affiliation{Multidisciplinary Centre for Advanced Research and Studies (MCARS), Jamia Millia Islamia, New Delhi 110025, India}
\affiliation{Astrophysics and Cosmology Research Unit,
	School of Mathematics, Statistics and Computer Science,
	University of KwaZulu-Natal, Private Bag X54001,
	Durban 4000, South Africa}
\date{\today}

\begin{abstract}
Low energy limits of a string theory suggests that the gravity action should include quadratic and higher-order 
curvature terms, in the form of dimensionally continued Gauss-Bonnet densities. Einstein-Gauss-Bonnet 
is a natural extension of the general relativity to higher dimensions in which the first and second-order
terms correspond, respectively, to general relativity and Einstein-Gauss-Bonnet gravity . We obtain five-dimensional
($5D$) black hole solutions, inspired by a noncommutative geometry, with a static spherically symmetric, 
Gaussian mass distribution as a source both in the general relativity and Einstein-Gauss-Bonnet gravity cases, 
and we also analyzes their thermodynamical properties. Owing the noncommutative corrected black hole, 
the thermodynamic quantities have also been modified, and phase transition is shown to be achievable. 
The phase transitions for the thermodynamic stability, in both the theories, are characterized by a 
discontinuity in the specific heat at $r_+=r_C$, with the stable (unstable) branch for $r < (>)\;  r_C$.  
The metric of the noncommutative inspired black holes smoothly goes over to the Boulware-Deser solution 
at large distance.  The paper has been appended with a calculation of black hole mass using holographic renormalization. 
\end{abstract}

\maketitle

\section{INTRODUCTION}
The Gauss-Bonnet term, which is the dimensionally extended version of the four-dimensional Euler density, 
is present in the low energy effective action of heterotic string theory \cite{30, 31, 32}, and 
it also appears in six-dimensional Calabi-Yau compactifications of \textit{M}-theory \cite{33}.
The Einstein-Gauss-Bonnet  theory is one of the natural generalization of Einstein's general relativity, 
introduced originally by Lanczos \cite{Lanczos}, and rediscovered by David Lovelock \cite{dll}. It  has some 
special characteristics among the larger class of general higher-curvature theories and this is an 
example of a theory with higher derivative terms. Nevertheless, the field equations are of second-order 
like in general relativity. The Einstein-Gauss-Bonnet theory, being a higher-dimensional member of 
Einstein's general relativity family, allow us to explore several conceptual issues of gravity in a 
broader setup and the theory is known to be free of ghosts about other exact backgrounds \cite{bd}. 
The theory represents a very interesting scenario to study how higher curvature corrections to black hole 
physics substantially change the qualitative features we know from our experience with black holes 
in general relativity. Since their inception, steady attention has been devoted to black hole solutions, 
including their formation, stability, and thermodynamics. The spherically symmetric static black hole  
solution for the Einstein-Gauss-Bonnet gravity was first obtained by Boulware and Deser \cite{bd,ms}, 
and later several authors explored exact black hole solutions including their thermodynamical properties 
\cite{egb,Cai:2001dz,Cai:2003gr,Ghosh,Sahabandu:2005ma,Cai:2003kt,Wiltshire:1988uq,Kastor:2006vw}. 
Several generalizations of the Boulware-Desser solution with matter source have also been  obtained 
\cite{som1,hr,sgr,sgsm,Ghosh:2014pga} and, from viewpoint of gravitational collapse to a black hole \cite{Dadhich:2013bya,Dadhich:2012cv, Jhingan:2010zz,Ghosh:2010jm}

The motivation to consider such a theory is that the Gauss-Bonnet term naturally appears as the
next-to-leading term in heterotic string effective action. On the other hand, the noncommutative geometry 
appears naturally from the study of open string theories. In particular, the noncommutative black holes 
are involved in the study of string and M-theory \cite{Haro}. It was shown that the gravitational wave signal GW150914  can be used
to place a bound on the scale of quantum fuzziness of noncommutative space-time \cite{Kobakhidze:2016cqh}.  The gravitational wave signal GW150914 was recently detected by LIGO and Virgo collaborations \cite{GW}.  It was shown  that the
leading noncommutative correction to the phase of the gravitational waves produced by a binary system
appears at the second order of the post-Newtonian expansion \cite{Kobakhidze:2016cqh}. Further,  the plausibility of using quantum mechanical transitions, induced by the combined effect of gravitational waves  and noncommutative  structure  to probe the spatial noncommutative  has been also reported \cite{Saha:2015fsa}.   The noncommutative spacetime  was 
originally introduced by Snyder \cite{snyder} to study the divergences in relativistic quantum field theory.    
The effects of noncommutative gravity can be made by formulating a model in which general relativity is 
its usual commutative form, but the noncommutative geometry leads to a smearing of matter distributions 
which is viewed as due to the intrinsic uncertainty embodied in the coordinate commutator of 
  \begin{equation}\label{nco}
  \left[x_a, x_b\right] = i \theta_{ab}, 
  \end{equation}
  where $\theta_{ab} $ is an anti-symmetric matrix which determines the fundamental cell discretization 
  of spacetime. 
This effective approach may be considered as an improvement to semiclassical gravity and a way to understand 
the noncommuative effects. Motivated by this idea, models of noncommutative geometry inspired 
Schwarzschild black holes were obtained by Nicolini, Smailagic, and Spallucci \cite{Nicolini:2005vd}, 
which was extended to the Reisnner-Nordstr{\"o}m model in four dimensions \cite{Ansoldi:2006vg}, generalized 
to higher-dimensional spacetime by Rizzo \cite{Rizzo:2006zb}, to charge in higher-dimensions 
\cite{Spallucci:2009zz}, and then to the BTZ black holes \cite{Kim:2007nx}. Further, recent 
years witnessed a significant interest in  noncommuative models 
\cite{Myung:2006mz,Kim:2008vi,Rahaman:2013gw,Banerjee:2008du} mainly due to its relevance in 
quantum gravity. A review of the noncommutative inspired model can be found in Ref.~\cite{Nicolini:2006}.
Thus, the main effect of noncommutativity is proposed to be the smearing out of conventional mass 
distributions. Hence, we will take, instead of the point mass, $M$, described by a $\delta$-function 
distribution, a static, spherically symmetric, Gaussian-smeared matter source, in $D-$dimensions 
\cite{Spallucci:2009zz,Rizzo:2006zb}, as
\begin{equation}\label{rho}
\rho_{\theta}(r) = \frac{\mu }{(4 \pi \theta)^{(D-1)/2}} e^{{-r^2}/{(4\theta)}}
\end{equation} 
The particle mass $ \mu $ diffused throughout a region of linear size $ \sqrt{\theta} $. Here $ \theta $ 
is the noncommutative parameter which is considered to be of a Planck length. 
Thus, (\ref{rho}) plays the role of a matter source and the mass is smeared around the region 
$ \sqrt{\theta} $ instead of locating at a point. This paper searches for a solution of 
five-dimensional ($5D$) Einstein-Gauss-Bonnet equations in the presence of a static, spherically 
symmetric Gaussian mass distribution to find a black hole solution inspired by the noncommutative geometry. 
Our starting point is to solve the full non-linear Einstein-Gauss-Bonnet equations for a static and 
spherically symmetric metric for a source (\ref{rho}). We study not only the structure of noncommutative  
solutions but also the thermodynamical stability of the system. In particular, we explicitly bring out how 
the effect of noncommutativity  can alter black hole solutions and their thermodynamical properties. 

The paper is organized as follows. In Sec.~\ref{NCEGB}, we find a general solution to the $5D$ 
spherically symmetric static Einstein-Gauss-Bonnet equations for the source ({\ref{rho}}), which allows 
us to discuss the problem of a noncommutative  inspired black hole  and also to discuss their 
thermodynamical properties. We also derive basic equations for $5D$ Einstein-Gauss-Bonnet gravity, which 
go over to the general relativity, for the case $\alpha=0$, and also discuss energy-momentum tensor for a 
noncommutative inspired  matter source. The thermodynamic properties of the solution derived in 
Sec.~\ref{NCEGB}, for the Einstein-Gauss-Bonnet gravity, is discussed in Sec.~\ref{EGB-thermodynamics}.  
A discussion on the thermodynamical stability of the noncommutative inspired  Einstein-Gauss-Bonnet 
black hole and phase transition is subject of Sec.~\ref{NCPT}. It ends with concluding remarks 
in Sec.~\ref{concluding}. The paper is appended with a thermodynamics of the noncommutative $5D$ Schwarzschild-Tangherilini  black hole. 

We use units which fix the speed of light and the gravitational constant via $8\pi G = c = 1$, and use 
the metric signature ($-,\;+,\;+,\;+,\;+$).

\section{Einstein-Gauss-Bonnet black holes}
\label{NCEGB}
The Gauss-Bonnet term is the only possibility for the leading correction to Einstein general relativity, 
for the slope expansion to be ghost free, in the low energy effective string theory 
\cite{bd,Wiltshire:1988uq}. The Einstein-Gauss-Bonnet gravity action \cite{egb,Ghosh:2014pga,Ghosh} 
in $5D$ can be written as
\begin{equation}\label{Action}
\mathcal{I}_{G}=\frac{1}{2}\int_{\mathcal{M}}dx^{5}\sqrt{-g}\left[  \mathcal{L}_{1} +\alpha \mathcal{L}_{GB}
 \right] + \mathcal{I}_{S},
\end{equation}
with $\kappa_5 =1$. $\mathcal{I}_{S}$ denotes the action associated with matter and $\alpha$ is a
coupling constant with dimension of $(\mbox{length})^2$  which is positive in the hetoretic string theory. 
The Einstein term is $ \mathcal{L}_1 = R$ and the second-order Gauss-Bonnet term $\mathcal{L}_{GB}$ is
\begin{equation}
\mathcal{L}_{GB}=R_{\mu\nu\gamma\delta}R^{\mu \nu\gamma\delta}-4R_{\mu\nu}R^{\mu\nu}+R^{2}.
\end{equation}
Here, $R_{\mu\nu}$, $R_{\mu\nu\gamma\delta}$, and $R$ are the Ricci tensor, Riemann tensor, and  
Ricci scalar, respectively. The variation of the action with respect to the metric $g_{\mu\nu}$ gives 
the Einstein-Gauss-Bonnet equations \cite{Cai:2003kt,Ghosh,Sahabandu:2005ma}
\begin{equation}\label{ee}
G_{\mu\nu}^{E}+\alpha G_{\mu\nu}^{GB}=T_{\mu\nu}^{S},
\end{equation}
where $G_{\mu\nu}^{E}$ is the Einstein tensor, while $G_{\mu\nu}^{GB}$ is explicitly given
by \cite{Kastor:2006vw}
\begin{eqnarray}
 G_{\mu\nu}^{GB} & = & 2\;\Big[ -R_{\mu\sigma\kappa\tau}R_{\quad\nu}^{\kappa
\tau\sigma}-2R_{\mu\rho\nu\sigma}R^{\rho\sigma}-2R_{\mu\sigma}R_{\ \nu
}^{\sigma} \nonumber \\ & &  +RR_{\mu\nu}\Big] -\frac{1}{2}\mathcal{L} _{GB}g_{\mu\nu}. 
\end{eqnarray}
We note that the divergence of Einstein-Gauss-Bonnet tensor $G_{\mu \nu}^{GB}$ vanishes. Here, we want to obtain noncommutative geometry inspired  $5D$ static, spherically symmetric solutions of Eq.~(\ref{ee}) and investigate its properties. We assume that the metric has a form \cite{Ghosh,Ghosh:2014pga,ms}
\begin{equation}\label{metric}
ds^2 = -f(r) dt^2+ \frac{1}{f(r)} dr^2 + r^2 \tilde{\gamma}_{ij}\; dx^i\; dx^j,
\end{equation}
where $ \tilde{\gamma}_{ij} $ is the metric of a $3D$ constant curvature space $k = -1,\; 0,\;$ or $1$. 
In this paper, we shall restrict to $k = 1$. The noncommutativity eliminates point like structure in 
 favor of smeared objects in the flat spacetime. The effect of smearing is mathematically 
implemented with a Gaussian distribution of minimal width $\sqrt{\theta}$
\cite{Nicolini:2005vd}.
We have assumed the form of the matter density $\rho$ above  and the condition $g_{00}= 1/g_{rr}$ 
(see Ref.~\cite{Spallucci:2009zz,Rizzo:2006zb} for further details), two components of the diagonal 
stress-energy tensor reads
\begin{eqnarray}\label{cemt}
T^0_0 &=& - T^r_r=  \rho_{\theta}(r) = \frac{\mu }{16 \pi^2 \theta^2} e^{{-r^2}/{(4\theta)}}. 
\end{eqnarray}
The Bianchi identity  $T^{ab};b=0$ \cite{Rizzo:2006zb} gives 
\begin{equation}
0= \partial_r T^r_r + \frac{1}{2} g^{00} \left[T^r_r-T^0_0\right] \partial_r g_{00} 
+ \frac{1}{2} \sum g^{ii}  \left[T^r_r-T^i_i\right] \partial_r g_{ii} 
\end{equation}
and the fact that $g^{ii} \partial_r g_{ii} = 2/r$, we obtain the energy momentum tensor in the $5D$ 
spacetime as 
\begin{eqnarray}\label{EMT}
T^t_t &=& T^r_r= \rho_{\theta}(r), 
\nonumber \\
T^{\theta}_{\theta} &=& T^{\phi}_{\phi} = T^{\psi}_{\psi} = \rho_{\theta}(r) + \frac{r}{3} 
\partial_r \rho_{\theta}(r), 
\end{eqnarray}
and the energy-momentum tensor  is completely specified by (\ref{EMT}). The static spherically symmetric 
black hole solution to the Einstein-Gauss-Bonnet Eq.(\ref{ee}), was first obtained by Boulware and 
Deser \cite{bd} to show the only spherically symmetric solution to the Einstein-Gauss-Bonnet gravity 
is Schwarzschild-Tangherilini type solution. Here, we find the noncommutative inspired 5D black hole solution 
in the Einstein-Gauss-Bonnet gravity. Interestingly, Eq.~(\ref{ee}) for the matter source (\ref{EMT}),  
admits a general solution 
\begin{equation} \label{sol:egb}
f_{\pm}(r) = 1+\frac{r^{2}}{4{\alpha}}\left[1\pm \sqrt{1+\frac{8\alpha \mu}{ r^4 \pi} 
\gamma\left(2, \frac{r^2}{4\theta}\right) }\right],
\end{equation}
by appropriately relating $\mu$ with  integrating constants $c_1$  \cite{hr} and 
$\gamma\left(2 \ , r^2/4\theta\, \right)$ is the lower incomplete Gamma function \cite{Spallucci:2009zz},
\begin{equation}
\gamma\left(2\ , r^2/4\theta\, \right)\equiv \int_0^{r^2/4\theta}\; du\; u\; e^{-u}.  
\end{equation}
In order to proceed further, we define the mass-energy $\mu(r)$ 
\begin{equation}\label{mass}
\mu(r) = \frac{2  \mu}{\pi} \gamma\left(2\ , r^2/4\theta\, \right),
\end{equation}
whereas the total mass-energy, M, measured by asymptotic observer \cite{Ansoldi:2006vg} is given by
\begin{equation}\label{mass1}
M= \lim_{r \rightarrow \infty} \mu(r) = \frac{2  \mu}{\pi} ,
\end{equation}
and solution (\ref{sol:egb}) becomes
\begin{equation} \label{sol:egb1}
f_{\pm}(r) = 1+\frac{r^{2}}{4{\alpha}}\left[1\pm \sqrt{1+\frac{8\alpha M}{r^{4}}
\gamma\left(2, \frac{r^2}{4\theta}\right) }\right],
\end{equation}
Eq.~(\ref{sol:egb1}) is an exact solution of the field equation (\ref{ee}) for matter source (\ref{EMT}), 
which in the limit  ${r/\sqrt{\theta}} \rightarrow \infty$ reduces to the Boulware and Deser \cite{bd,ms}Gauss-Bonnet black hole solution, and the negative branch of solution (\ref{sol:egb1}) reduces to that of noncommutative $5D$  \cite{Spallucci:2009zz,Rizzo:2006zb}. 
Whereas asymptotically far away, we have $\rho_{\theta}(r)=0$.   
\begin{figure*}
	\begin{center}
\begin{tabular}{c c }
\includegraphics[width=0.55\linewidth]{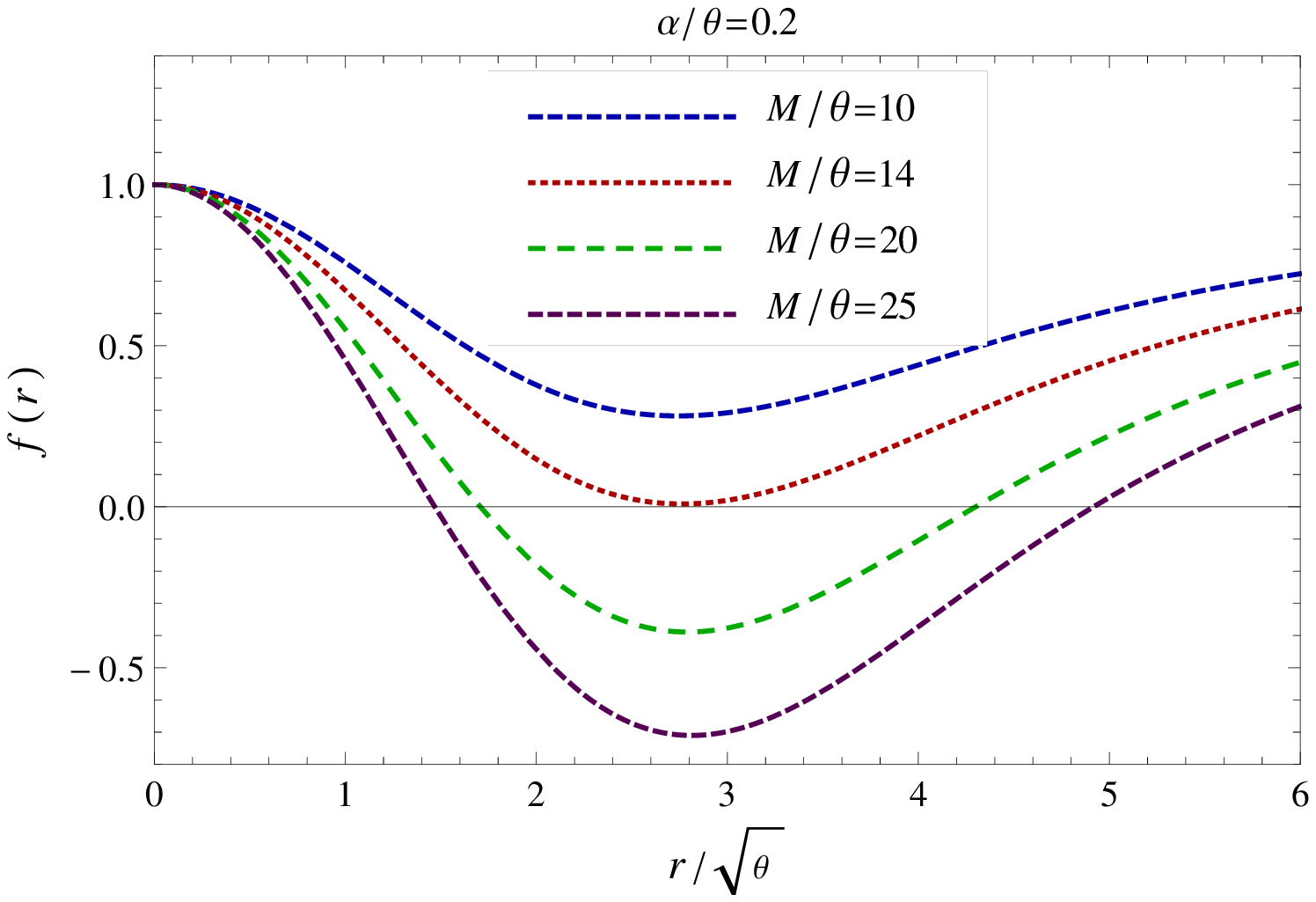}
\includegraphics[width=0.55\linewidth]{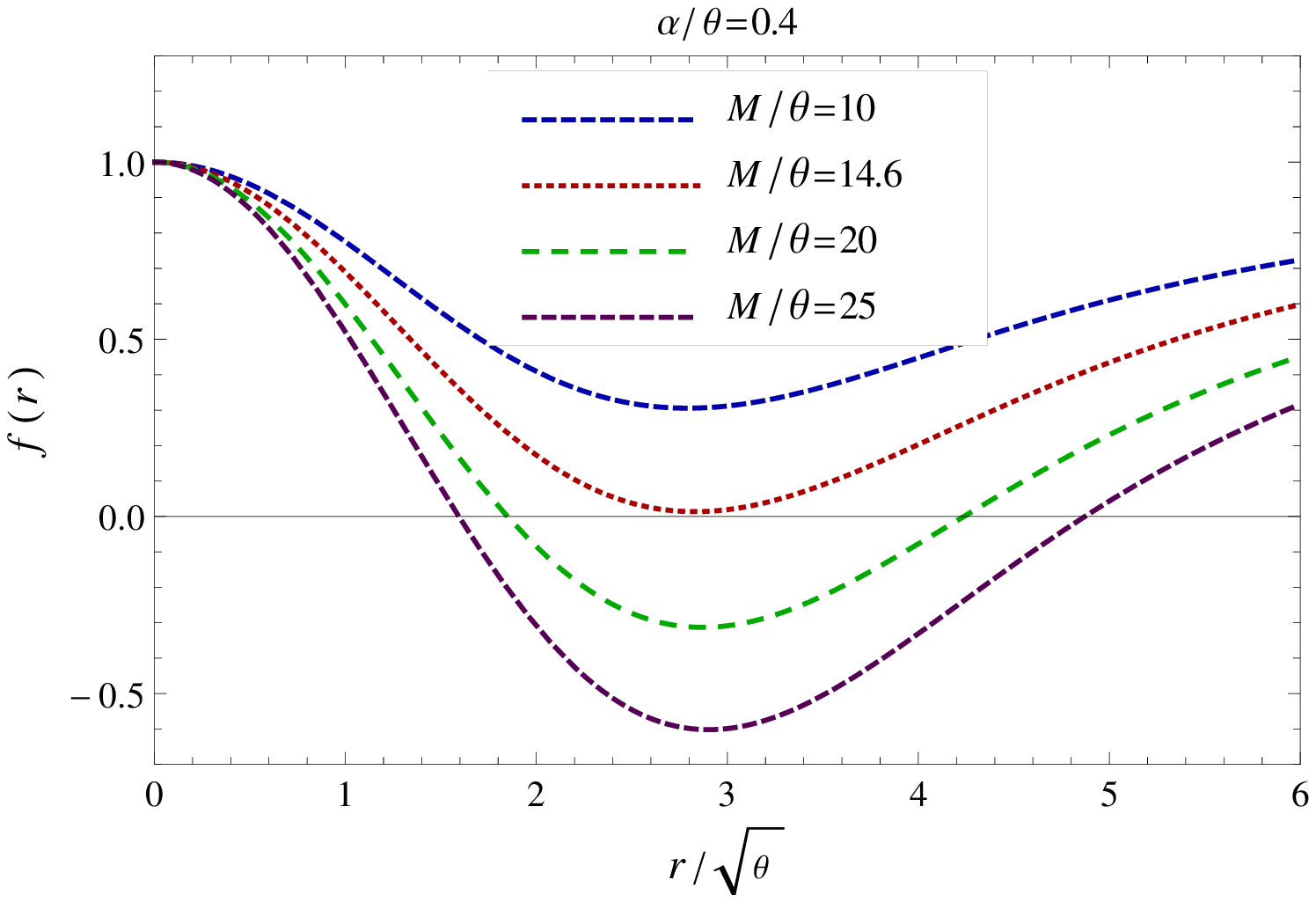}
\end{tabular}
\caption{\label{f5d} Plot of metric function $f(r)$ vs $r/\sqrt{\theta}$, for various values of 
$M/\theta$ for the $5D$ noncommutative inspired Einstein-Gauss-Bonnet black hole.}
	\end{center}
\end{figure*}
By definition, $r=r_+ $ is a horizon when $g^{rr}(r_+)=0 \, \mbox{or} \,  f(r_+)=0$, which imply
\begin{equation}\label{eh}
	r_{+} = \sqrt{M \gamma\left(2, \frac{r_+^2}{4\theta}\right)-2 \alpha}.  
\end{equation}

In the long distance limit, the effect of the noncommutativity can be neglected, and in the short distance, 
significant changes are expected due to noncommutativity. Equation~(\ref{eh}) cannot be solved analytically 
and hence it is shown in the Fig.~\ref{f5d}, by plotting $f(r)$ as a function of $r$. The intersection 
with $ r $-axis gives the location of the horizons. A plot of $f(r)=0$, indicates where it might dip below 
zero. Figure~\ref{f5d} shows that there will be a range of parameters for which there is no black hole, 
and that the simplest black hole cases will generically have an inner and outer horizon,
the two cases separated by an extreme black hole with degenerate horizons. Thus, the effect of 
noncommutativity leads to an additional horizon, since the commutative Einstein-Gauss-Bonnet black holes 
have just one horizon \cite{Ghosh:2014pga}. It turns out that for a given $\alpha$, there exists a critical 
value of mass  $M$, $M_C$, and critical horizon radius $r_{+}$, $r_C$, such that $g^{rr}(r_C)=f(r_C)=0$ 
has a double root $r_C$, which corresponds to an extremal black hole with degenerate horizons. 
When $M>M_C$, $f(r)=0$ has two simple zeros, and has no zeros for $M<M_C$ (cf. Fig.~\ref{f5d} and 
Table~\ref{tes}). These two cases corresponds, respectively, to a nonextremal black hole with 
two horizons viz., a Cauchy horizon (CH) and an event horizon (EH), and no black hole.   
It is worthwhile to mention that the critical values of $M_C$ and $r_C$ are $\alpha$ dependent, e.g., for
$\alpha=0.2,\;0.4$, respectively $M_C=14 \sqrt{\theta},\; 14.6 \sqrt{\theta}$ and
$r_C=2.76182,2.88023$ (cf. Fig.~\ref{f5d}). Indeed, both $M_C$ and $r_C$
increases with the increase in $\alpha$.  Note that the outer horizon has always a radius larger 
than the critical radius. In the limit $\alpha \rightarrow 0$ and $r/\sqrt{\theta} \rightarrow \infty$, 
Eq.~(\ref{eh}) gives $5D$ Schwarzschild-Tangherilini event horizon $r_+^2=M$. We observe that when 
$r/\sqrt{\theta} \rightarrow \infty$, the black hole horizon is located at $r_{+} = \sqrt{M -2 \alpha}$ 
and that $0<M<\alpha$, we do not have naked singularity \cite{Ghosh,Ghosh:2014pga}.  
\begin{figure*}
\begin{tabular}{c c}
\includegraphics[width=0.55\linewidth]{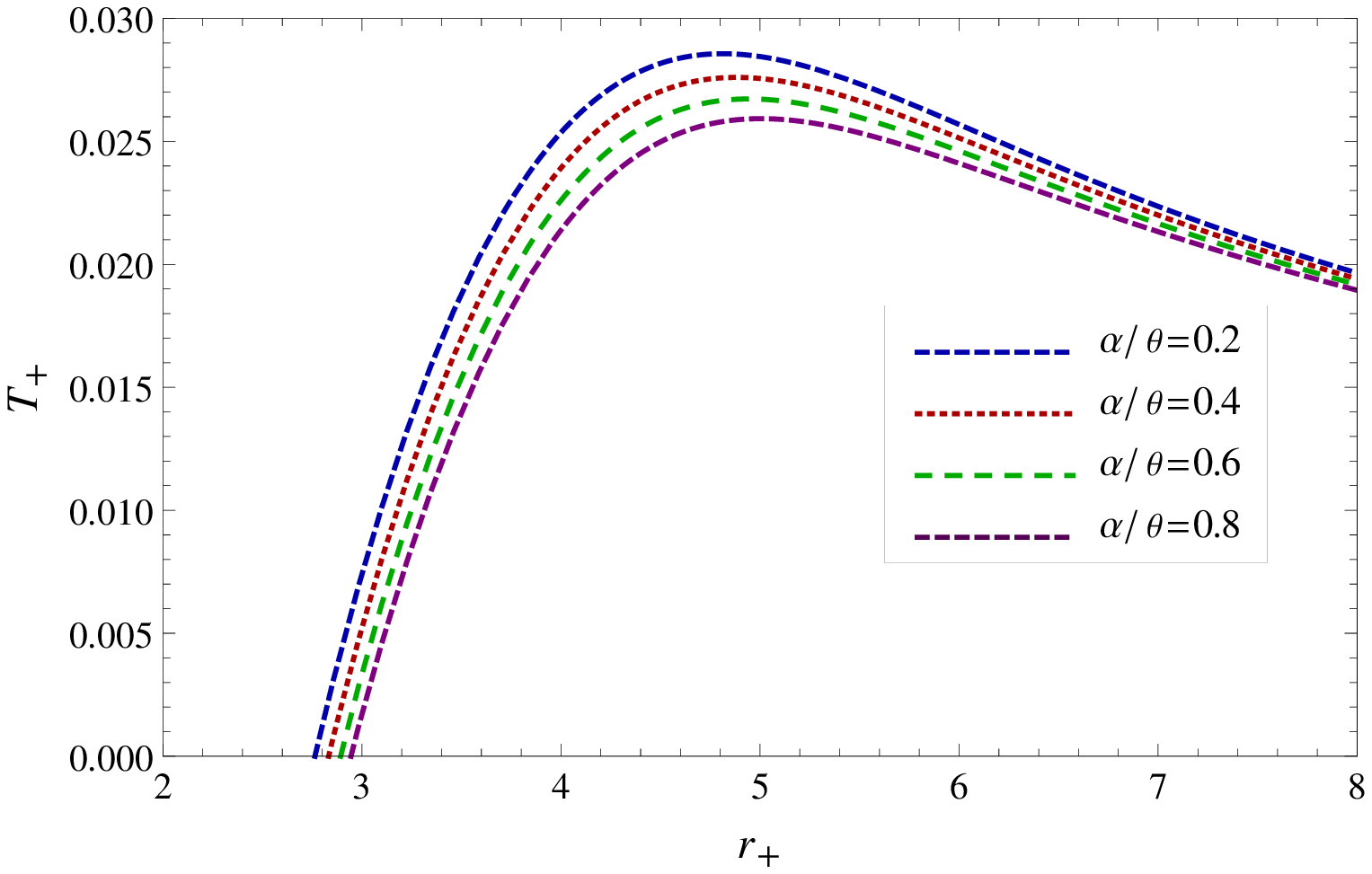}
\includegraphics[width=0.55\linewidth]{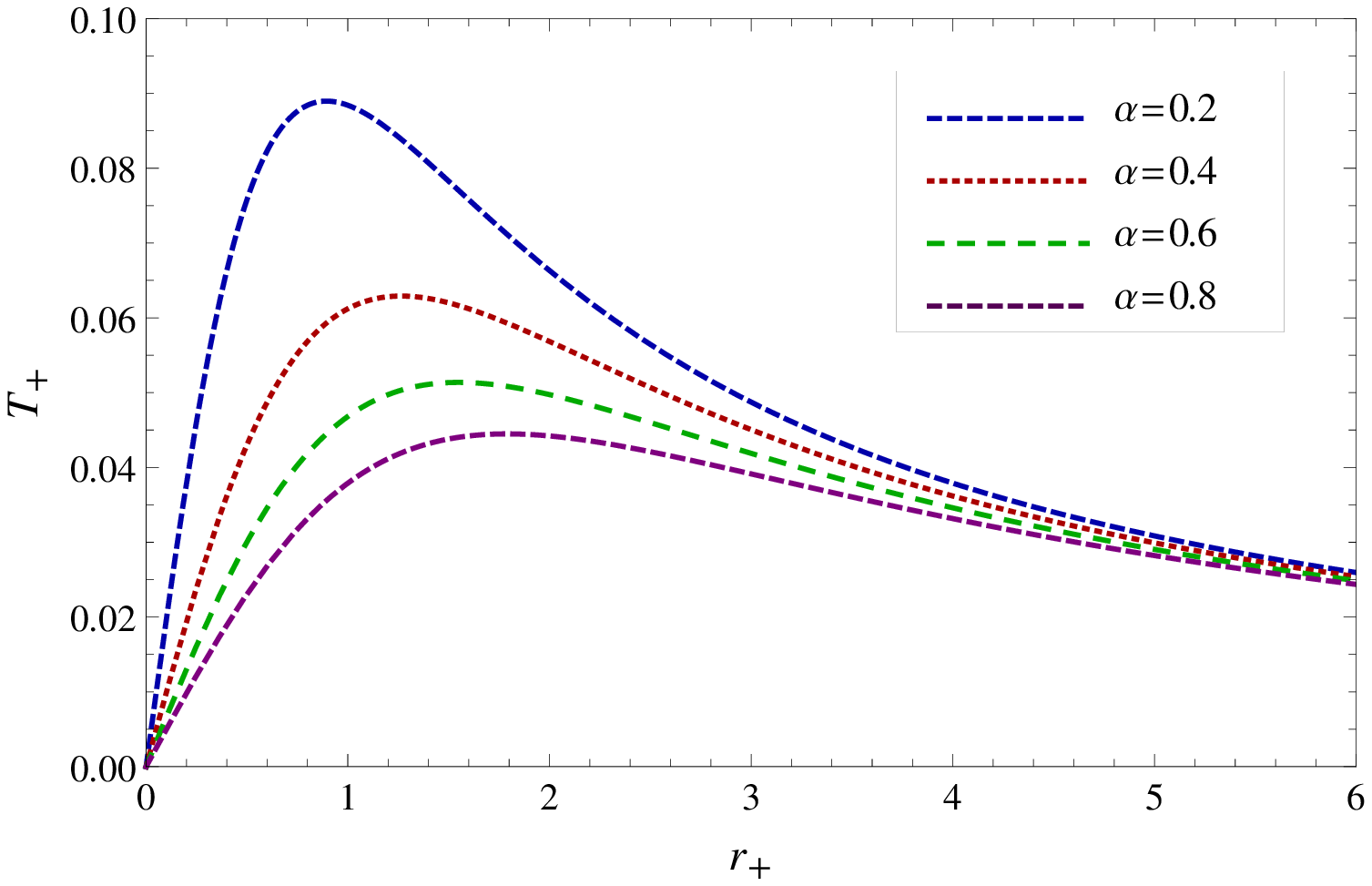}
\end{tabular}
\caption{\label{fig:egb:T} The Hawking temperature ($ T_+ $) vs horizon radius $r_+$ for different values of $\alpha$, in $\sqrt{\theta}$ units,  for the noncommutative $5D$ Einstein-Gauss-Bonnet black hole.  $T_+=0$ for $r_+=r_0=2.76248\sqrt{\theta},\; 2.83194\sqrt{\theta},\;  2.89142 \sqrt{\theta}$ and $2.94361\sqrt{\theta}$, respectively for $\alpha=0.2\;\theta,\;0.4\;\theta,\;0.6\;\theta,\; \mbox{and}\; 0.8\; \theta, $ which is compared with the commutative counterpart. }
\end{figure*}
Thus, the metric (\ref{metric}) with (\ref{sol:egb1}) describes a noncommutative inspired 
Einstein-Gauss-Bonnet black hole and it may be important to understand how the noncommutativity affects 
the thermodynamical properties including the stability. 
\section{Black hole thermodynamics}
\label{EGB-thermodynamics}
In this section we shall discuss and reckon the main thermodynamical properties of the $5D$ noncommutative 
inspired Einstein-Gauss-Bonnet black holes, and henceforth discussion shall be restricted  to the negative 
branch of the solution (\ref{sol:egb1}). The  black holes  are characterized by their mass $(M_+)$. 
From Eq.~(\ref{sol:egb1}), mass of the black hole  can be expressed in terms of its horizon radius 
($r_+$) as
\begin{eqnarray}\label{M1}
	M_{+} =  \frac{ r_{+}^2+2\alpha}{\gamma\left(2, \frac{r_{+}^2}{4\theta}\right)}. 
\end{eqnarray}
Equation~(\ref{M1}) reduces to the $5D$ Schwarzschild-Tangherilini black hole with mass $M_{+}=r_+^2 $, when 
$\alpha = 0$ and $r/\sqrt{\theta} \rightarrow \infty$. The Hawking temperature associated with the 
black hole is defined by $T=\kappa/2\pi$, where $\kappa$ is the surface gravity 
\cite{Sahabandu:2005ma,Ghosh:2014pga} defined by
\begin{equation}
\kappa^{2}=-\frac{1}{4}g^{tt}g^{ij}g_{tt,i}\;g_{tt,j},
\end{equation}
which on inserting the metric function (\ref{sol:egb1}) becomes
\begin{equation}
\kappa=\left\vert \frac{1}{2}f^{\prime}(r_+)  \right\vert.
\end{equation}
Accordingly, the modified Hawking temperature of the black hole (\ref{sol:egb1}) on the outer horizon 
$r_+$ , reads
\begin{eqnarray}\label{temp1}
T_+  &=&  \frac{1}{4 \pi(r_+^2 + 4 \alpha)} \left[{2r_{+}}- (r_+^2 + 2 \alpha)
\frac{\gamma' \left(2, \frac{r_{+}^2}{4\theta}\right)}{\gamma\left(2, \frac{r_{+}^2}{4\theta}\right)}\right].  
\end{eqnarray}
Equation~(\ref{temp1}) gives the modified Gauss-Bonnet black hole temperature, and in the limit 
$r/\sqrt{\theta}\rightarrow0$, we recover the commutative Gauss-Bonnet black hole temperature as
\begin{equation}
T_{+}^{\mbox{EGB}} =\frac{1}{4 \pi} \left[\frac{2r_{+}}{r_+^2 + 4 \alpha} \right], 
\end{equation} 
exactly same as obtained in \cite{Ghosh:2014pga,Sahabandu:2005ma,ms}. Here $T_+=1/(2\pi r_+)$ is the 
Hawking temperature of the $5D$ Schwarzschild-Tangherilini black hole \cite{Ghosh}. However, in the 
Einstein-Gauss-Bonnet gravity, it remains finite \cite{Ghosh:2014pga,ms}, and the same is  for the 
noncommutative case as shown in Fig.~\ref{fig:egb:T}. Also, when  $\alpha/\theta \neq 0$, the 
Hawking temperature ($T^+$) has a peak which decreases and shifts as $\alpha/\theta $ increase 
(cf. Fig.~\ref{fig:egb:T}). The maximal Hawking temperature ($T_+^{Max}$), in $\sqrt{\theta}$ units, 
occurs at a critical radius $r_C^T$, is depicted in Table~\ref{tes1}. It turns out that the $T_+^{Max}$ 
decreases with $\alpha$, while $r_C^T$ increases.   Therefore, at the initial stage of Hawking radiation, the black hole temperature increases
as the horizon radius decreases.  The temperature,  as is shown in Fig.~\ref{fig:egb:T},  grows during its evaporation until it reaches the maximum value  and then falls down to zero at the extremal black hole.

Next, we calculate black hole  entropy, which can be obtained by the first law of thermodynamics. 
The entropy of a black hole in general relativity satisfies the area law that state, i.e., the entropy 
of a black hole is a quarter of the event horizon area \cite{Cai:2003kt,Sahabandu:2005ma,ms}. 
The black hole behaves as a thermodynamical system and hence, the quantities associated with it must 
obey the first law of thermodynamics \cite{Cai:2003kt,Sahabandu:2005ma,Ghosh:2014pga}
\begin{equation}\label{flaw}
dM_+ = T_{+}dS_{+}.
\end{equation}
Hence, the entropy can be obtained from the integration 
\begin{eqnarray}\label{ent:formula}
S_{+} = \int {T_+^{-1} dM} = \int {T_+^{-1}\frac{\partial M_+}{\partial r_+} dr_+}, 
\end{eqnarray}
and substituting (\ref{M1}) and (\ref{temp1}) into (\ref{ent:formula}), the entropy of the 
noncommutative inspired Einstein-Gauss-Bonnet gravity black holes, determined by using 
Eq.~(\ref{ent:formula}), reads
\begin{eqnarray}\label{S:EGB}
S_{+} = 4\pi\int  \frac{4\alpha+r_{+}^2}{\gamma\left(2, \frac{r_{+}^2}{4\theta}\right)}d r_+.
\end{eqnarray}
The entropy for our model differs from the expression of the general relativity in the limit 
$r/\sqrt{\theta}\rightarrow 0$, the entropy (\ref{S:EGB}) in Einstein-Gauss-Bonnet also reduces to  \begin{equation}\label{S1}
S_{+} = 4\pi \left(4 \alpha r_+ + \frac{r_+^3}{3}\right),
\end{equation}
exactly same as in the Ref. \cite{Ghosh:2014pga,Spallucci:2009zz,ms}. Notice that $S_{+} = 4\pi  {r^3}/{3}$ is the entropy of the $5D$ Schwarzschild-Tangherilini  black hole \cite{Ghosh}.

\section{Thermodynamic Stability and phase transition}\label{NCPT}
\begin{figure*}
	\begin{tabular}{c c}
		\includegraphics[width=0.55\linewidth]{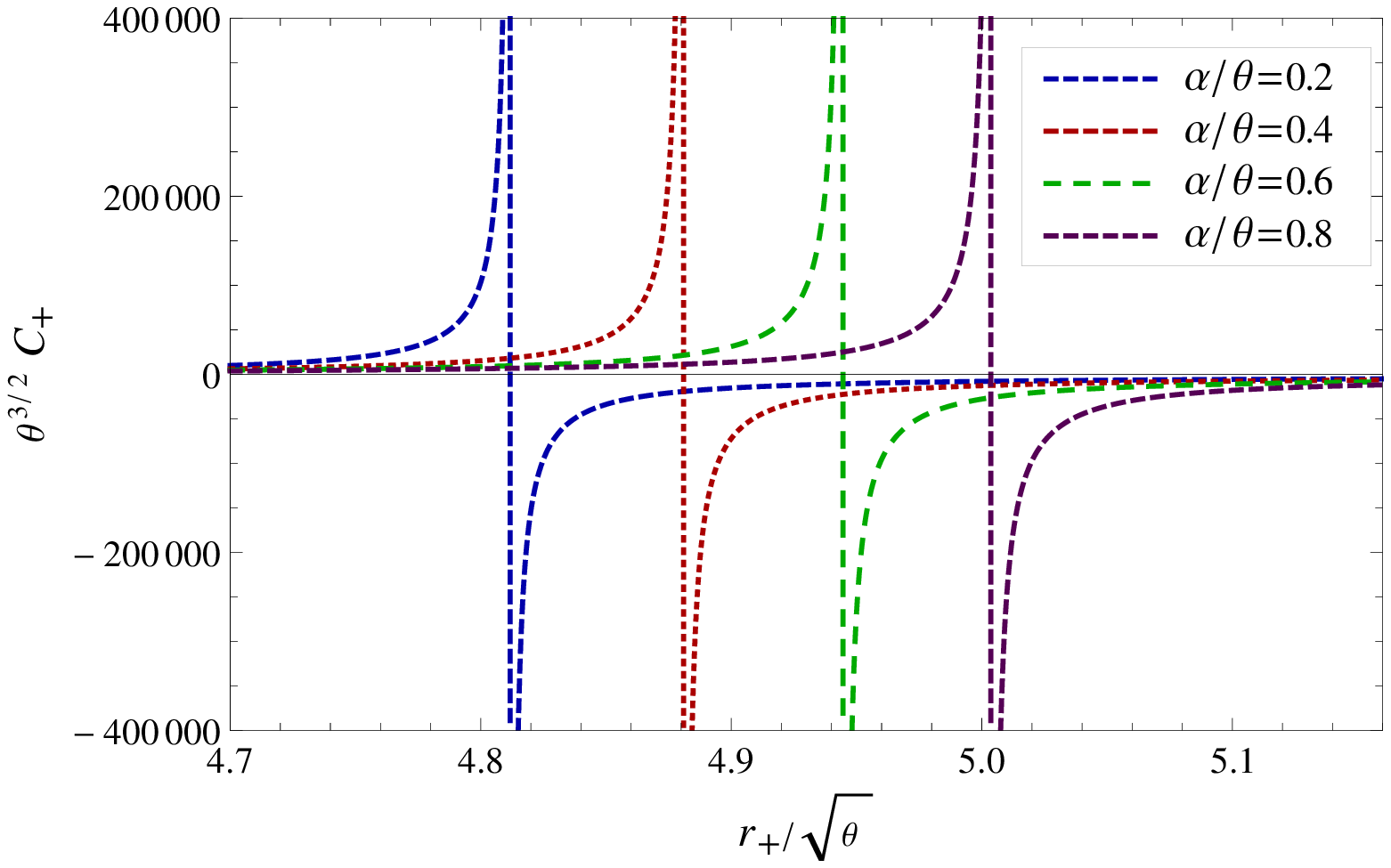}
		\includegraphics[width=0.55\linewidth]{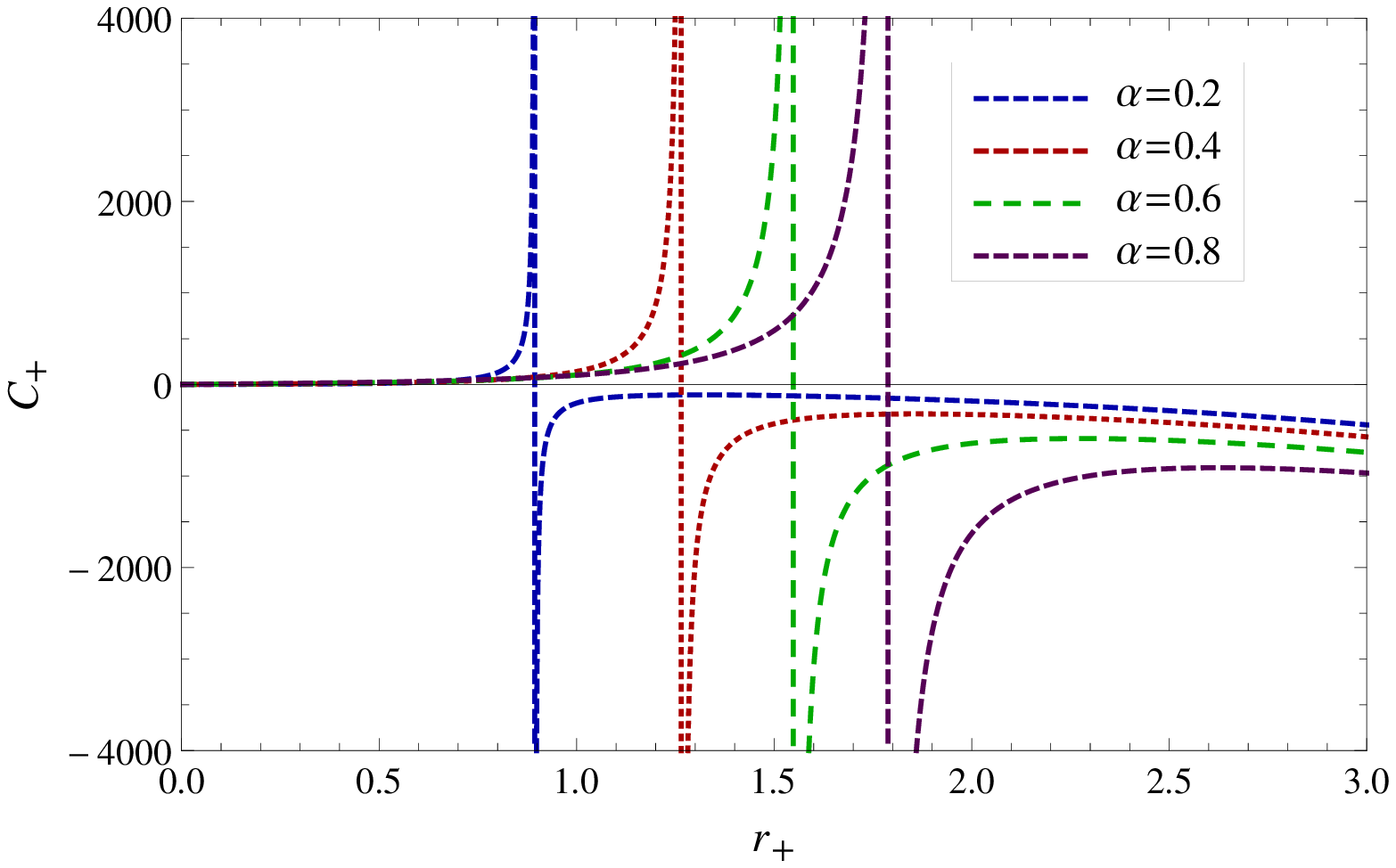}
	\end{tabular}
	\caption{\label{fig:egb:sh} The specific heat  $ (\theta^{3/2}\; C_+)$  vs horizon radius ($r_+/\sqrt{\theta}$) for the $ 5D $ Einstein-Gauss-Bonnet black hole for different values of  $\alpha$, which is compared with the commutative counterpart.}
\end{figure*}

Finally, we analyze how the noncommutativity influences the thermodynamic stability of the 
Einstein-Gauss-Bonnet black holes. The  thermodynamical stability of a black hole is performed by study 
of its heat capacity. The heat capacity of the black hole is defined as  in \cite{Cai:2003kt} 
\begin{equation}\label{sh_formula}
C_+ = \frac{\partial{M_+}}{\partial{T_+}}= \left(\frac{\partial{M_+}}{\partial{r_+}}\right)
\left(\frac{\partial{r_+}}{\partial{T_+}}\right). 
\end{equation} 
On using Eqs.~ (\ref{M1}) and (\ref{temp1}), yields
\begin{widetext}
\begin{eqnarray} \label{egb:sh}
&& C_+ = \frac{4 \pi (r^2+4\alpha)^2 \left[(r^2+2 \alpha) \gamma' - 2 r \gamma\right]}{(r^2+4\alpha) 
(r^2+2\alpha)\gamma \gamma{''} - (r^4 +6\alpha r^2+8 \alpha^2) \gamma'^2 + 4 \alpha r\gamma \gamma' + 
2(r^2-4\alpha)\gamma^2}, \nonumber \\ && \mbox{with}\, \gamma = \gamma\left(2, \frac{r_{+}^2}{4\theta}\right)
\end{eqnarray}
\end{widetext}
again the large distance limit leads to 
\begin{equation}\label{cegb}
C_{+}^{EGB} = - 4 \pi r_+ \frac{(r_{+}^2 +4 \alpha)^2}{(r_{+}^2-4 \alpha)},
\end{equation}
which is exactly same as the commutative Einstein-Gauss-Bonnet case \cite{Ghosh:2014pga}. 
It is well known that the thermodynamic stability of the system is related to the sign of the heat capacity 
($C_+$); if it is positive ($ C_+ >0$), then the black hole is stable; when it's negative ($ C_+ <0$), 
 the black hole is said to be unstable. It is difficult to analyze the heat capacity analytically,
hence we plot it in Fig.~\ref{fig:egb:sh} for different values of $\alpha$. The heat capacity is positive 
for $r_+<r_C$ and thereby suggesting the thermodynamical stability of the black hole. On the other hand, 
the black hole is unstable for $r_+>r_C$. The heat capacity is discontinuous at $r_+=r_C$ means the 
second-order phase transition happens there \cite{davies}. Interestingly, the discontinuity of the heat 
capacity occurs exactly at $r_C^T$, where the Hawking temperature attains a maximum value  of the black hole 
mass increases with increasing $r_+$. Hence, the phase transition occurs from a lower mass black hole 
with the positive heat capacity to a higher mass black hole with negative heat capacity.  
It may be noted that the critical radius $r_C$ changes drastically due to noncommutativity, 
thereby affecting the thermodynamical stability. Further, the critical value $r_C$ is sensitive to the 
Gauss-Bonnet parameter $\alpha$ (cf. Fig.~\ref{fig:egb:sh}), and the critical parameter $r_C$ also 
increases with $\alpha$. 

\section{Discussion}\label{concluding}
The Einstein-Gauss-Bonnet gravity has a number of additional nice properties than the general relativity which is not enjoyed by other higher-curvature theories.   It has been widely studied because it can be obtained in the low energy limit of string theory. On the other hand, motivated by string theory arguments \cite{snyder}, noncommutative geometry has been explored extensively with an important emphasis on the black hole spacetime.  One can assert that noncommutative structure of the space-time is one of the exotic outcomes of the string theory, and the paradigm of noncommutative geometry is even more general. The noncommutative geometry provides an effective framework to study short-distance space-time dynamics. Here, we have obtained $5D$ static spherically symmetric black hole solutions to general relativity and Einstein-Gauss-Bonnet gravity inspired by noncommutative geometry at a short distance. We have found a $5D$ Boulware-Deser-like metric which reproduces exactly the $5D$ Boulware-Deser and $5D$ Schwarzschild-Tangherilini solution in the appropriate limits.  We characterized the new solution by calculating the horizons which could be at least two, as against one in the $5D$ commutative,  describing a variety of charged, self-gravitating objects, including an extremal black hole with degenerate horizons and a non-extremal black hole with Cauchy and event horizons.  It is seen that the new solution smoothly interpolating between a de Sitter core around the origin and an ordinary  Boulware-Deser like metric at large distance.  

We have also analyzed the black thermodynamical quantities like the  black hole mass, Hawking temperature, entropy, specific heat and in turn also analyzed the thermodynamical  stability of black holes. It turns out that due to noncommutativity, the thermodynamical quantities also  get corrected.  The Hawking temperature does not diverge as the event horizon shrinks down instead it reaches a maximum value for a critical radius and then drops down to zero.   The entropy of a black hole, in general relativity, obeys the area law, but not for  noncommutative black holes.  Regarding the thermodynamic stability, we showed that the heat capacity can be positive or negative depending on the horizon radius. In particular, the phase transition is characterized by the divergence  of specific heat at a critical radius $r_C$ which changes with the Gauss-Bonnet parameter $\alpha$.  These black holes are thermodynamically stable, in both the theories, with a positive heat capacity for  the range $0 < r < r_C$ and unstable for $r>r_C$. It would be important to understand how these black holes  with positive specific heat ($C_+>0$) would emerge from thermal radiation through a phase transition.  The heat capacity becomes singular at a critical radius $r_C$ which corresponds to the maximum Hawaking temperature. It turns out that the specific heat $C_+0$ goes from being infinitely negative to infinitely positive and then down to a finite positive in the analogy of the Hawking-Page phase transition in the AdS black hole. The infinite change at $r_C$ indicates a thermodynamic behavior at the extremal black hole. We also observe that a thermodynamically unstable region ($C_+>0$) appears for $r>r_C$. The critical radius is $r_C$  is also affected by the noncommutativity. 

The results presented here are the generalization of the previous discussions, on the $5D$ black hole,  in general relativity and Einstein-Gauss-Bonnet gravity, in a more general setting, and the possibility of a further generalization of these results the Lovelock gravity is an interesting problem for future research.

\acknowledgements
S.G.G. would like to thank SERB-DST Research Project Grant No. SB/S2/HEP-008/2014 and DST INDO-SA bilateral project DST/INT/South Africa/P-06/2016 and also to IUCAA, Pune for the hospitality while this work was being done. Special thanks to M. Amir for help in plots and Rahul Kumar for fruitful discussions.   

\appendix*
\section{I Exact solutions for noncommutative geometry $5D$ Schwarzschild-Tangherilini  black hole ($ \alpha=0 $)}
\label{sol-GR}
The  metric that describes  $5D$ noncommutative geometry inspired Schwarzschild-Tangherilini black hole reads
\begin{eqnarray}\label{metric5d1}
& & ds^2 = -\left[1-\frac{2 \mu}{r^2\pi}  \gamma\left(2, \frac{r^2}{4\theta}\right)\right] dt^2+ \frac{1}{  \left[1-\frac{2 \mu}{r^2\pi}  \gamma\left(2, \frac{r^2}{4\theta}\right)\right]} dr^2 \nonumber \\ &  & + r^2 d\omega^2_3,
\end{eqnarray}
with $d\omega^2_3$ is metric on the 3-sphere. 
 Using  mass-energy $\mu(r)$ and total mass-energy, M, measured by asymptotic observer defined above, 
the metric, in terms of $M$, yields 
\begin{eqnarray}\label{metric5d}
&& ds^2 = -\left[1-\frac{M}{r^2}  \gamma\left(2, \frac{r^2}{4\theta}\right)\right] dt^2+ \frac{1}{  \left[1-\frac{M}{r^2}  \gamma\left(2, \frac{r^2}{4\theta}\right)\right]} dr^2 \nonumber \\ & & + r^2 d\omega^2_3.
\end{eqnarray}
In the limit $r/\sqrt{\theta}\to\infty $, Eq.~(\ref{metric}) reduces to the $5D$ Schwarzschild-Tangherilini 
metric. 
\begin{figure*}
	\begin{tabular}{ c c }
		\includegraphics[width=0.75\linewidth]{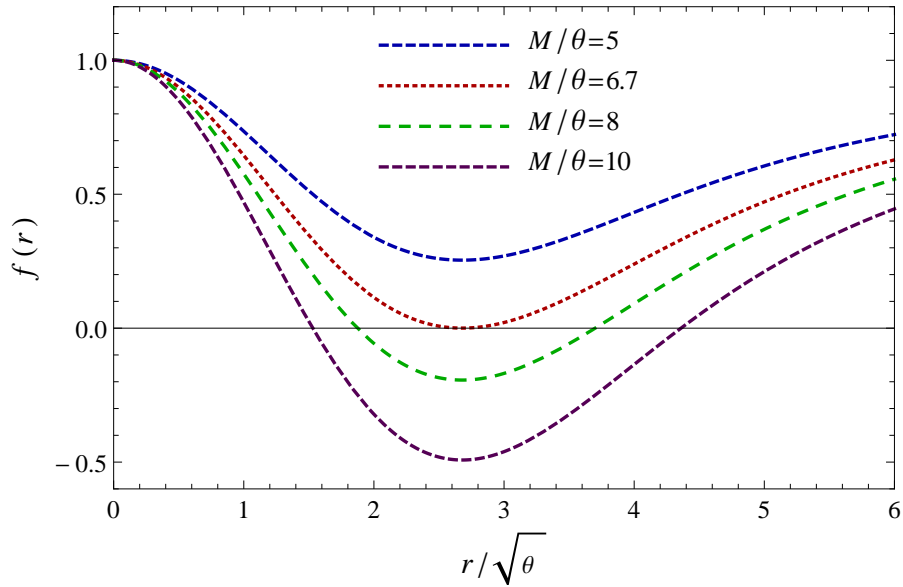}
	\end{tabular}
	\caption{\label{5dc1} Plot of metric function $f(r)$ vs $r/\sqrt{\theta}$, for various values of $M/\theta$ for $5D$ noncommutative inspired black hole.}
\end{figure*}
The event horizon $r_+$ of the black hole (cf. Fig.~\ref{5dc1}), satisfy $g_{tt}(r_+)=0$, i.e.,
\begin{equation}
r_{+}^2 = {M} \gamma\left(2, \frac{r^2}{4\theta}\right),
\end{equation}
\begin{figure*}
	\begin{tabular}{ c c }
		\includegraphics[width=0.65\linewidth]{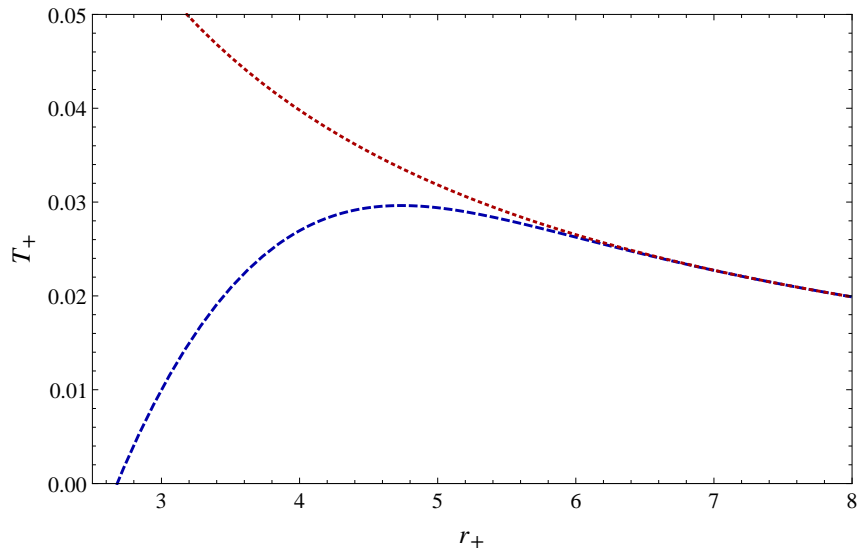}
	\end{tabular}
	\caption{\label{5dtemp} The temperature ($T_+$) vs horizon radius $r_+$ for nocommutative $5D$ Schwarzschild-Tangherilini  black hole, and commutative counterpart temperature.}
\end{figure*}
which in the limit, $r/\sqrt{\theta} \rightarrow 0$, we recover $5D$ Schwarzschild-Tangherilini event horizon $r_+^2=M$.  The gravitational mass of a black hole is determined  as
\begin{eqnarray}\label{mass}
M_{+} = \frac{r_{+}^2}{ \gamma\left(2, \frac{r^2}{4\theta}\right)}.
\end{eqnarray}
Equation~(\ref{mass}) takes the form of the $5D$ Schwarzschild-Tangherilini black hole mass $M=r_+^2$ \cite{Ghosh}. The Hawking temperature associated with the black hole on the outer horizon $r_+$, reads
\begin{eqnarray}
T_{+} = \frac{\kappa}{2\pi}=\frac{1}{4 \pi r_{+}} \left[2 -  \frac{r_+ \gamma{'}\left(2, \frac{r_+^2}{4\theta}\right)}{\gamma \left(2, \frac{r_+^2}{4\theta}\right)}\right]. \label{temp}
\end{eqnarray}
Then, we can easily see that the temperature is positive as shown in Fig.~\ref{5dtemp}. Taking the limit $r/\sqrt{\theta}\rightarrow \infty$, we recover the temperature for $5D$ general relativity \cite{Sahabandu:2005ma,Ghosh:2014pga}:
\begin{equation}\label{tem}
T_{+} = \frac{1}{2\pi r_+},
\end{equation}
which shows that Hawking temperature diverges as $r_+ \rightarrow 0$ (cf. Fig.~\ref{5dtemp}).
Next, we turn to calculate the entropy associated with the black hole horizon which in $4D$ obeys the 
area formula. The black hole behaves as a thermodynamic system. Hence, quantities associated with it 
must obey the first law of thermodynamics \cite{Cai:2003kt}, $dM_+ = T_{+}dS_{+}$,
\begin{eqnarray}\label{S:GR}
S_{+} =  \int \frac{\pi r_+^2}{\gamma\left(2, \frac{r_+^2}{4\theta}\right)} dr.
\end{eqnarray}
In the limit $\alpha \rightarrow 0$, we arrive at  
\begin{eqnarray}\label{S:GR}
S_{+} = \frac{4\pi r_{+}^{3}}{3}. 
\end{eqnarray}
Thus, we note the quantity $4\pi r_{+}^3 /3$ of Eq.~(\ref{S:GR}) is just area of the black hole horizon, 
i.e., we conclude that the five-dimensional black hole also obeys an area law. In proper units, when $r/\sqrt{\theta}\rightarrow \infty$, the Eq.~(\ref{S:GR}) may be written as $S_{+}=A/4G$.  
\begin{figure*}
	\begin{tabular}{ c c }
		\includegraphics[width=0.65\linewidth]{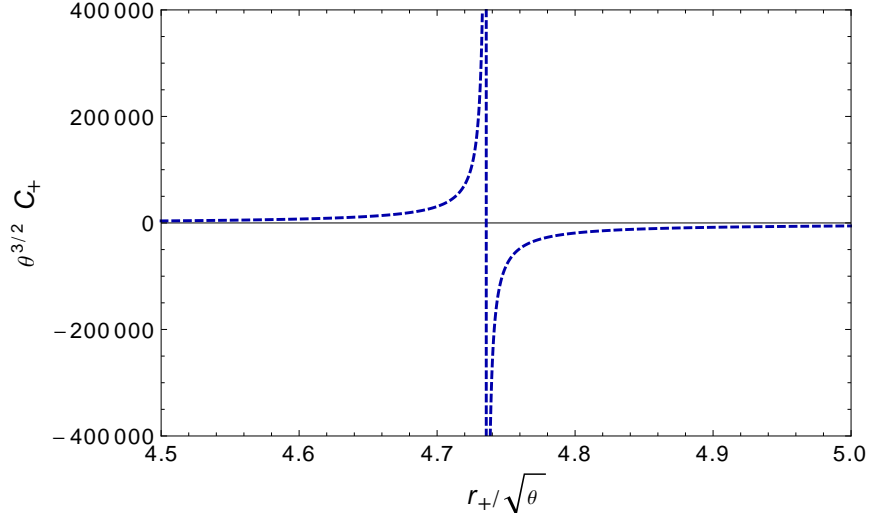}
	\end{tabular}
	\caption{\label{sh5d} The specific heat ($C_+$) vs horizon radius $r_+$ for the $5D$ nocommutative $5D$ Schwarzschild-Tangherilini  black hole. The black hole thermodynamically unstable when $C_+<0$, and stable when $C_+>0.$}
\end{figure*}
The heat capacity reads
\begin{equation}\label{sh_formula}
C_{+}= \frac{4 \pi r_{+}^3 \left[r \gamma{'}\left(2, \frac{r_+^2}{4\theta}\right) 
- 2 \gamma\left(2, \frac{r_+^2}{4\theta}\right)\right]}{2 \gamma^2\left(2, \frac{r_+^2}{4\theta}\right)
- r_{+}^2 \left[\gamma{'}^2 \left(2, \frac{r_+^2}{4\theta}\right)
- \gamma\left(2, \frac{r_+^2}{4\theta}\right) \gamma{''} \left(2, \frac{r_+^2}{4\theta}\right) \right]} 
\end{equation}
which is plotted inn the Fig~.\ref{sh5d} and, in the limit $r/\sqrt{\theta}\rightarrow \infty$, reduces to
\begin{equation}
C_{+} = -4 \pi r_{+}^{3}.
\end{equation}
Thus, the $5D$ Schwarzschild-Tangherlini always have negative heat capacity indicating thermodynamic 
instability of the black hole. 

\appendix*
\section{II Mass and temperature using renormalization method}
 In this section, we follow the quasilocal formalism to study the mass from holographic renormalization in asymptotically flat spacetime. The noncommutative inspired 5D black hole solution in Einstein-Gauss-Bonnet theory is given by the line element  (\ref{metric}) with metric function $f(r)$ (\ref{sol:egb1}). In this method, we enclose the given region of spacetime with some "quasilocal surface", which can be easily extend upto spatial infinity. It must be noted that gravitational action diverges due to integration over infinite region of spacetime. One way to regularize the action is quasilocal formalism. In order to regulate the calculation, we have to put a cut-off on the spacetime at large $r=r_0$, which is a finite value of radial coordinate. We denote the resulting spacetime by $\mathcal{M}_0$ with a boundary $\partial\mathcal{M}_0$ defined by metric $h_{ij}$. The stress tensor could be renormalized under the scheme of holographic renormalization by adding some appropriate counter term (defined on the boundary $\partial\mathcal{M}_0$) in the action \cite{Batrachenko:2004fd}.
The action is thus given as
\begin{eqnarray}
\mathcal{I}=\mathcal{I}_G-\frac{1}{8\pi G}\int_{\mathcal{\partial M}_0}d^4x\sqrt{-h}\Theta,
\end{eqnarray}
where, $\mathcal{I}_G$ is action in Einstein-Gauss-Bonnet gravity defined in Eq.~(\ref{Action}), $\Theta$ is trace of extrinsic curvature tensor $\Theta_{ij}$ at boundary. The corresponding stress-tensor
can be obtained by varying the action as follow
\begin{eqnarray}
\tau_{ij}=\frac{2}{\sqrt{-h}}\frac{\delta \mathcal{I}}{\delta h_{ij}}.
\end{eqnarray}
Due to the presence of time translation symmetry, the associated conserved mass reads as
\begin{equation}
M=\oint_{\Sigma} d^3y\sqrt{\sigma}n^i \tau_{ij}\xi_{[t]}^j.
\end{equation}
where $\Sigma$ is hypersurface at the boundary described by the metric $\sigma_{ij}$, $n^i$ is the normal vector to hypersurface and $\xi^i$ is the Killing vector of the symmetric group.
Following \cite{Batrachenko:2004fd}, we have 
\begin{eqnarray}
\tau_{tt}=\frac{1}{8\pi G}\left(\Theta_{tt}-h_{tt}\Theta\right),
\end{eqnarray}
where
\begin{equation}
\Theta_{tt}=-h_{tt}\sqrt{f(r)}\left(\frac{1}{2f(r)}\frac{df(r)}{dr}\right),\quad
\Theta=-\sqrt{f(r)}\left(\frac{1}{2f(r)}\frac{df(r)}{dr}+\frac{3}{r}\right),
\end{equation}
The spacelike hypersurface $\Sigma$ is a constant-$t$ surface at boundary $\mathcal{M}_0$, whose unit normal could be easily calculated through the definition
\begin{equation}
n_{\alpha}=\frac{-\partial_{\alpha}t}{|h^{ij}\partial_{i}t\partial_{j}t|^{1/2}}=-\frac{1}{\sqrt{f(r)}}.
\end{equation}
Thus conserved mass is written as
\begin{equation}
M= -\frac{1}{8\pi G}\int_{r\rightarrow \infty} (r^3 \sin^2\theta\sin\phi)\left(-\frac{1}{\sqrt{f(r)}}\right)\left(\frac{3 f(r)^{3/2}}{r}\right)d\theta d\phi d\psi.
\end{equation}
The physical noncommutative inspired black hole in Einstein-Gauss-Bonnet gravity is described by the negative branch of solution Eq.~(\ref{sol:egb}), which at large $r$ limit behaves as
\begin{eqnarray}
f(r)&=&1+\frac{r^2}{4\alpha}\left[1-\sqrt{1+\frac{8\alpha\mu}{\pi r^4}\gamma\left(2,\frac{r^2}{4\theta}\right)}\right],\nonumber\\
&\approx& 1-\frac{r^2}{4\alpha}\left[\frac{1}{2}\left(\frac{8\alpha\mu}{\pi r^4}\right)-\frac{1}{8}\left(\frac{8\alpha\mu}{\pi r^4}\right)^2+.....\right].
\end{eqnarray}
Therefore, the conserved mass reads
\begin{equation}
M= \frac{3\pi}{4 G}\left(-r^2+\frac{\mu}{\pi}+\frac{2\mu^2\alpha}{\pi^2 r^4}+ \cdots \right).
\end{equation}
If we extend the spacetime boundary ($\partial\mathcal{M}_0$) upto spatial infinity ($r\to \infty$), then under renormalisation process we can leave the first term and finally we get
\begin{equation}
M=\frac{3\pi}{8G}\left(\frac{2\mu}{\pi}\right)=\frac{3\pi}{8G}m,
\end{equation}
where $m$ as defined in Eq.~(\ref{mass1}).
Which is corrected mass for 5D noncommutative inspired Einstein-Gauss-Bonnet black hole.

Next to estimate the Hawking's temperature associated with 5D noncommutative inspired EGB black hole, we use the Euclidean coordinate and calculate the  periodicity of time. The corresponding horizon temperature is inverse of this period \cite{Astefanesei:2009wi}. With ($t\rightarrow \iota\tau$), the line element (\ref{metric}) becomes 
\begin{equation}
ds^2=f(r)d{\tau}^2+f(r)^{-1}dr^2+r^2d\Omega_{3}^2.
\end{equation}
On using the transformation $\mathcal{X}=\sqrt{f(r)}$, we obtain
\begin{equation}
ds^2=\frac{4}{(df(r)/{dr})^2}\Big(d{\mathcal{X}}^2+\frac{(df(r)/{dr})^2}{4} {\mathcal{X}}^2d{\tau}^2\Big). 
\end{equation}
Hence, near the horizon ($r=r_+$), we notice the singularity, to remove it we must identify the time period with 
\begin{equation}
\frac{1}{2}\frac{df(r)}{dr}\Delta\tau=2\pi\Rightarrow \Delta\tau=\frac{4\pi}{df(r)/dr}. 
\end{equation}
Therefore, the black hole horizon temperature reads
\begin{equation}
T =\frac{1}{\Delta \tau}=\frac{1}{4\pi} \frac{df(r)}{dr},
\end{equation}
which is exactly the same as obtain in Sec. \ref{EGB-thermodynamics}.

\end{document}